# Archaeoastronomy of the "Sun path" at Borobudur

Giulio Magli
School of Architecture, Urban Planning and Construction Engineering,
Politecnico di Milano, Italy
Giulio.Magli@polimi.it

*The world-famous Javanese temple of Borobudur is located at the westernmost end of a straight line which includes two "satellite" temples, Mendut and Pawon. Originally, the three buildings were probably connected by a processional path running along this line. It is shown here that the alignment points, at the horizon, where the sun sets in the days of the zenith passages. This orientation is likely intentional and related to a ritual procession which connected the three temples, mimicking the path of the sun in the sky.*

## 1. Introduction

The Javanese temple of Borobudur (Fig. 1) is one of the greatest Buddhist monuments in the world, built between the 8th and 9th century during the Syailendra Dynasty (Mcguigan 1995). The monument may be described as a giant "stupa" hill built on five concentric square terraces. The walls and balustrades are decorated with fine low reliefs, and around the circular platforms are 72 small chapels in stupa form, each containing a Buddha statue. The temple is endowed with two "satellite" temples, Mendut and Pawon, probably built in the same period (Chandra 1980).

A number of pages have been written about the symbolism embodied in the details of the construction and in the numbers devised in the Borobodur architectural features, allegedly connected with astronomical cycles. In many cases, this kind of "numerology" arises by chance and has little or nothing to do with the true intentions of the builders of monuments of this kind, which were rather (or mostly) meant as explicit, easily readable symbols of the religion on which the temporal power relied its rights. A paradigmatic case in this context in Asia is that of the Angkor state temples, very close both temporally and ideologically to Borobodur; however - at Angkor exactly as at Borobodur - the fact that esoteric astronomical "numerology" is typically a modern invention does not mean that the builders were not interested in the celestial cycles and in connecting their monuments with the sky, and modern Archaeoastronomy is the science which allows us to approach with rigorous methods this kind of issues (Magli 2017).

In the present paper we apply such methods to the complex composed by Borobodur and by its two "satellite" temples. It is very well known indeed that the temples are located along a straight line: from "east to west" (the orientation is *not* cardinal, as we shall specify below) we find Mendut, Pawon at 1150 meters as the crow flies, and finally, further 1750 m west, Borobudur (Fig. 2). This topographical relationship looks hardly casual, as probably a processional way ran along the line in ancient times (the path is still partly mimicked by the modern road from Borobodur to Pawon up to the river Pogo). The orientation of the axis could be chosen at will by placing the satellite temples in an appropriate way, and the macroscopic deviation from the east-west direction appears rather odd at the first sight, since the Borobodur main complex is instead oriented to the cardinal points. It is, therefore, worth considering the possibility of an astronomically significant orientation.

## 2. Archaeoastronomy of the Borobodur axis

To study the orientation of the Borobodur axis we can use satellite imagery as the area is very well covered by Google Earth. In what follows we consider as azimuth of the axis that of the line connecting the summit of Pawon with the summit of Borobodur, as it is natural to do since the main temple was with all probability the final destination. This line crosses the perimeter of the smaller Mendut temple, with an error of a few meters with respect to its summit (Fig. 2). The value of the azimuth is 263°, thus with a deviation of 7° from the east-west line; the horizon corresponding to this line of sight has an altitude of 1° 48' yielding a absolute declination of 7° 8', while the opposite azimuth (83°) has horizon height 2° 10' yielding a declination 6° 41' (declinations are calculated using the program Getdec by Clive Ruggles).
An astronomical interpretation of the axis has already been attempted (Long 2009, Long and Voute 2008). These authors propose an orientation to the rising of the star Altair, of the constellation Aquila (The Eagle). Effectively, in 800 AD Altair had a declination 6° 25' ; however, also other brilliant stars had declinations comparable to 6° 41', such as Betelgeuse (6° 19') and Procyon (7° 31'), and the cultural arguments given by these authors to justify the choice of Altair appear rather weak. The same authors also suggest a solar interpretation (it is not specified if they think that the stellar alignment was intended together with a solar one or if one of the two occurred by chance). They mention the zenith passage sunset as a possibility, but get involved in a rather complicated discussion about the apparent height of Borobodur from different positions – and consequently the height at which the sun is seen to set on different days; a few astronomy errors and an exaggerate attention to the accuracy of the data further plague the discussion. Overall, although the merit of a first discussion of the astronomy of the Borobudur axis undoubtedly goes to Long and Voute, it is apparent that the topic needs a full reassessment using the methods of scientific Archaeoastronomy.
Actually, it appears that the declination of the Pawon-Mendut-Borobodur direction – which was also the direction of processional events - is the most interesting from the archaeoastronomical point of view. It is indeed very close to minus the latitude of the place (7° 8' vs. 7° 36' south). This means that the axis points quite precisely to the setting of the sun on the days of the zenith passages (11 October and 28 February). As it is easy to check, the Borobodur summit *does not* emerge from the horizon when it is seen from the beginning of the axis, so the zenith sun was (and is) seen to set on the hills behind the temple (Fig. 3). In other words, this is a astronomical alignment which was – or at least, could have been – established independently from Borobodur construction. The error committed is less than a solar diameter (30') and it must be taken into account that the trajectory of the sun at these latitudes is very steep so that very precise measurements have not to be expected. For the very same reason, however, the effect of the sun setting behind Borobodur, and finally inside it when the distance decreases, can be enjoined already when the sun is relatively high. The very same phenomenon has been documented by the present author (Magli 2017) at the Bayon temple in Angkor on the equinoxes; interestingly enough, the Bayon presents similarities with Borobodur, in that it is a Buddhist temple (built around the end of the 12 century) oriented to the cardinal points and literally crowded on all sides by stone images of benevolent Buddha (and/or of the king represented as such).

Of course, the possibility of a pure chance for the zenith alignment of the axis cannot be completely excluded; however, a true east-west orientation would have been more coherent with the orientation of Borobodur proper without creating logistic difficulties in the project (along the west-east direction from Borobodur there is a small bent in the Pogo river which – if belonging to the original course, a thing which is very doubtful – would have implied at most a slight deviation - a couple of degrees - to the south of east in order to avoid crossing the river two times). To strengthen the point of a astronomical orientation it is worth considering also the orientation of the other two temples. The axis of Pawon is ~289° (some existing data giving a value two degrees less do not seem compatible with satellite images); with an horizon height of 35' this yields a declination of 18° 40'. The axis of Mendut is ~301°, the horizon is nearly flat and this yields a declination 30° 47'. It is obvious therefore that Pawon is in the solar range (dates around mid May-end of July) while Mendut is not. Existing explanations awoke the span of the constellation Gemini at the time of construction: the two, distant temples would allegedly point to the "extremal points" of the setting of this constellation (Long and Voute 2008). Clearly this is quite a far-fetched explanation, but some kind of explanation must exist since these orientations are uncommon for the region and there is no topographical constraint for them. Actually, there is a sound possibility: the Moon. As is well known, in the course of a 18,6 years cycle the maximal declination of the Moon in her monthly cycle undergoes a slow variation from a minimum to a maximum, equal to the obliquity of the ecliptic $\varepsilon$ minus/plus the obliquity of the earth-moon plane ($\iota=5° 9'$) with respect to the ecliptic. This leads to a minor standstill at declination $\varepsilon-\iota$ and a maximal standstill at declination $\varepsilon+\iota$. In 800 AD the obliquity of the ecliptic was about 9' greater than today so $\varepsilon=23° 39'$ and the two standstills correspond to declinations 28° 48' and to 18° 30' respectively. The last matches impressively well the orientation of Pawon, while the first is not far (less than two degrees in declination, corresponding to less than 2 degrees also in azimuth) from that of Mendut (parallax corrections are negligible at these latitudes).

Since the minor standstill of the Moon is always mimicked by the sun two times a year, it is impossible to distinguish it from a solar orientation in the case of a single building. However, the coincidence of two buildings possibly related to the two standstills is, to say the last, impressive. It is worth mentioning that the role of the moon is quite relevant in Buddhism, since festivals and recurrences associated with Buddha's life are timed by the full moon. In this respect it is important to remember that precise azimuths for the major standstills of the Moon are very difficult to individuate, and major standstills lunar orientations should always be understood as aimed to the full moon closest to the solstice, which always attains a declination close to the extremal one in the years of the standstills. The choice of orientation to the extrema of the moon might thus have arisen from calendrical reasons.

## 3. Discussion

The three temples at Borobodur belong to the Mahayana Buddhism. The details of the cults practised are unsure, but a relationship certainly existed between the temples and the proclaimed divine nature of the kings who ordered their construction. In this connection, a possible, symbolic relationship between the three monuments was investigated in details by Moens (1951). In this controversial but anyhow scholarly work, the idea is that the temples were connected by a "magical birth" ritual, in which the monarch's consecration occurred both as the Buddha and as King. Moens proposed a ritual based on an analogy with the sun path in the sky in one day, and thus endowed with three main "stations":  east, zenith, and west. To these steps corresponded for the west, the beginning of the western staircase at Borobudur; for the east,  Mendut; and for the zenith, Pawon.
Although this interpretation is well known, it has never been referred explicitly to the specific days of the zenith passages, a connection which instead looks natural:  if the "solar path" ritual had to be referenced into in the architecture of the temples, and if the zenith culmination of the sun was, as it seems, a fundamental ingredient of the ritual, then we would expect  the procession to go in the direction from sunrise to sunset, and the processional path to be oriented in such a way as to indicate the zenith sunset, as it actually occurs.
Finding comparison belonging to the same cultural context would also be of help, but one the problems is that Borobodur architectural conception is almost unique. Besides the already mentioned Bayon, as far as the present author is aware the unique, vaguely reasonable comparison is the so-called  108 stupas monument, located on a hillside directly on the western bank of the Yellow River at Qingtongxia, Ningxia, China. The monument is slightly later than Borobudur, as it was constructed  during the Western Xia dynasty (1038–1227 AD), as part of a greater Buddhist temple complex. It is composed by 108 stupas of sun-dried mud bricks, arranged in rows disposed in a triangular formation which narrows with height, from 19 stupas on the first row to the uppermost single one. A front view of this monument  is actually quite reminiscent of one side of Borobodur.

As far as the present author is aware, the orientation of the 108 stupas monument has never been studied. The azimuth is 120° which, with an horizon height close to zero, gives a impressive declination -24° that is, very close to the winter solstice sunrise. The monument is therefore, with hardly any doubt, astronomically oriented although not to the same solar phenomenon of the Borobodur axis; of course however, at the latitude of Ningxia about 37° north, zenith passages do not occur.

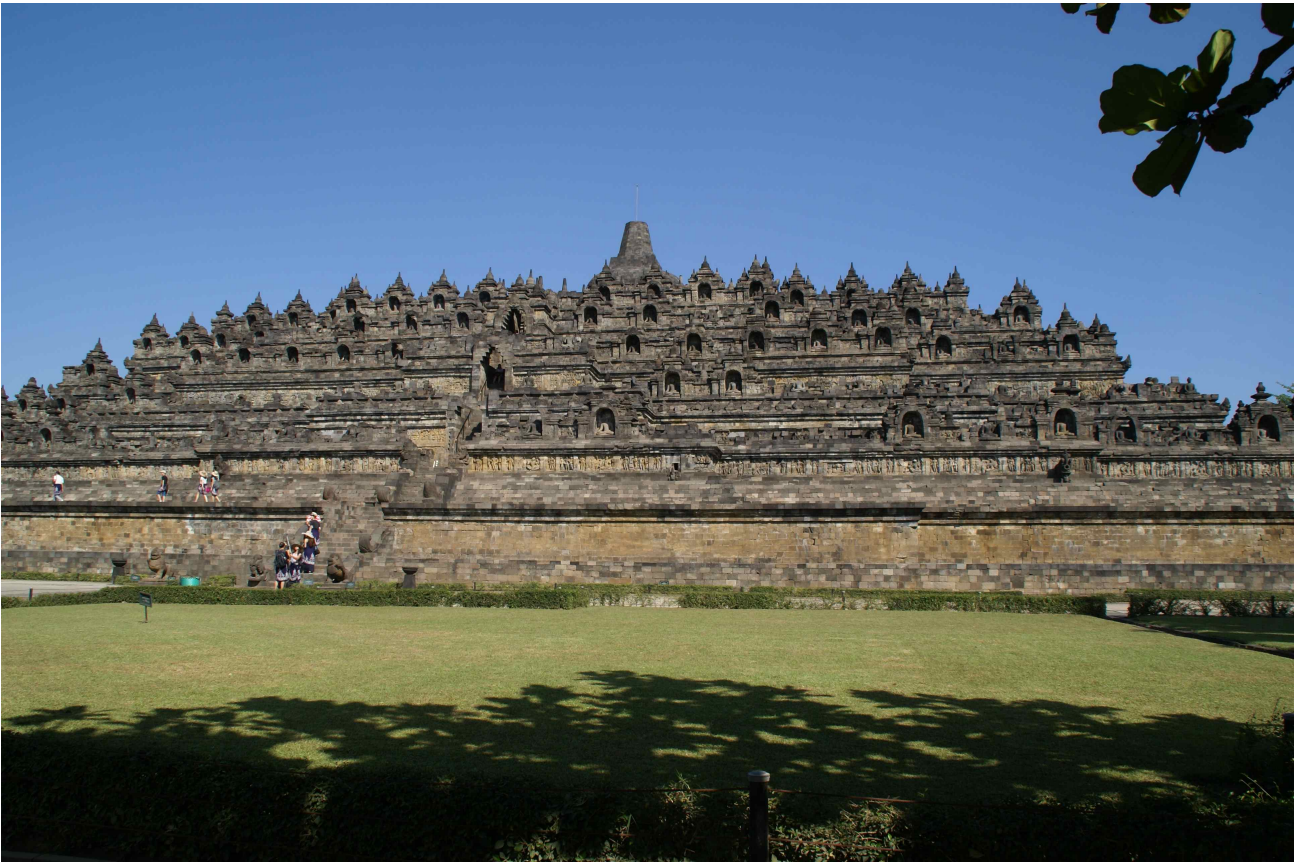

**Fig. 1 The temple of Borobodur**

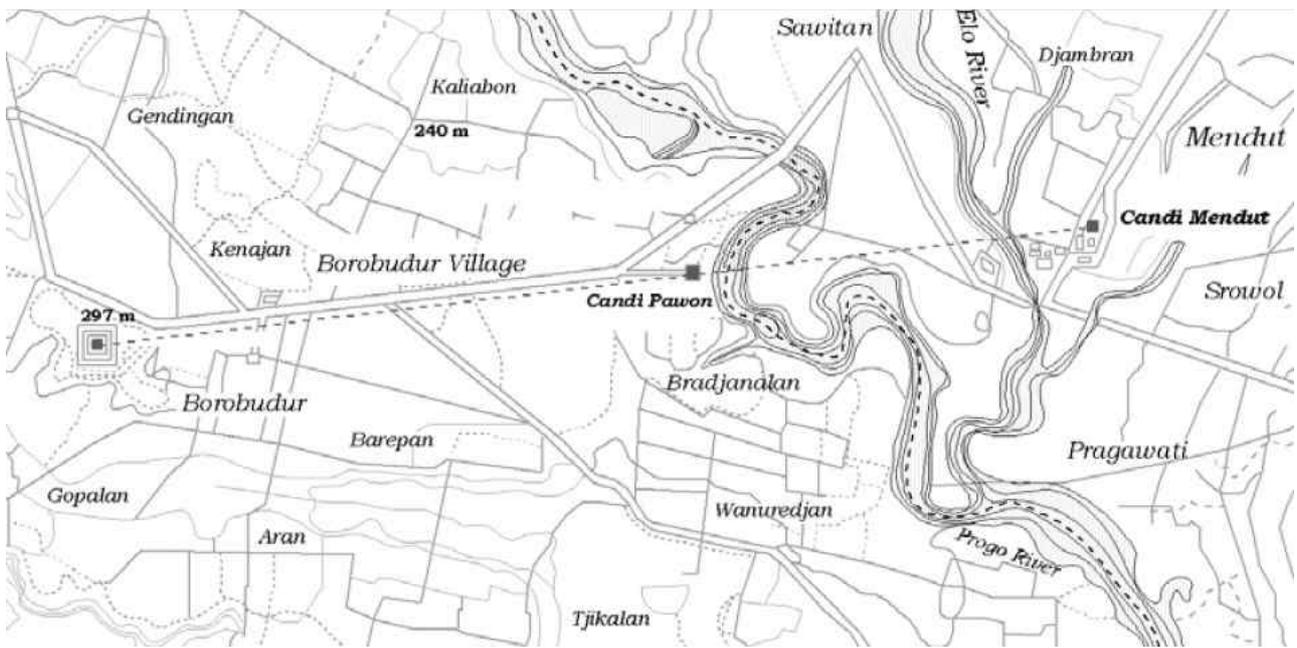

**Fig.2 The axis connecting Borobodur, Pawon and Mendut
(adapted from Moens 1951).**

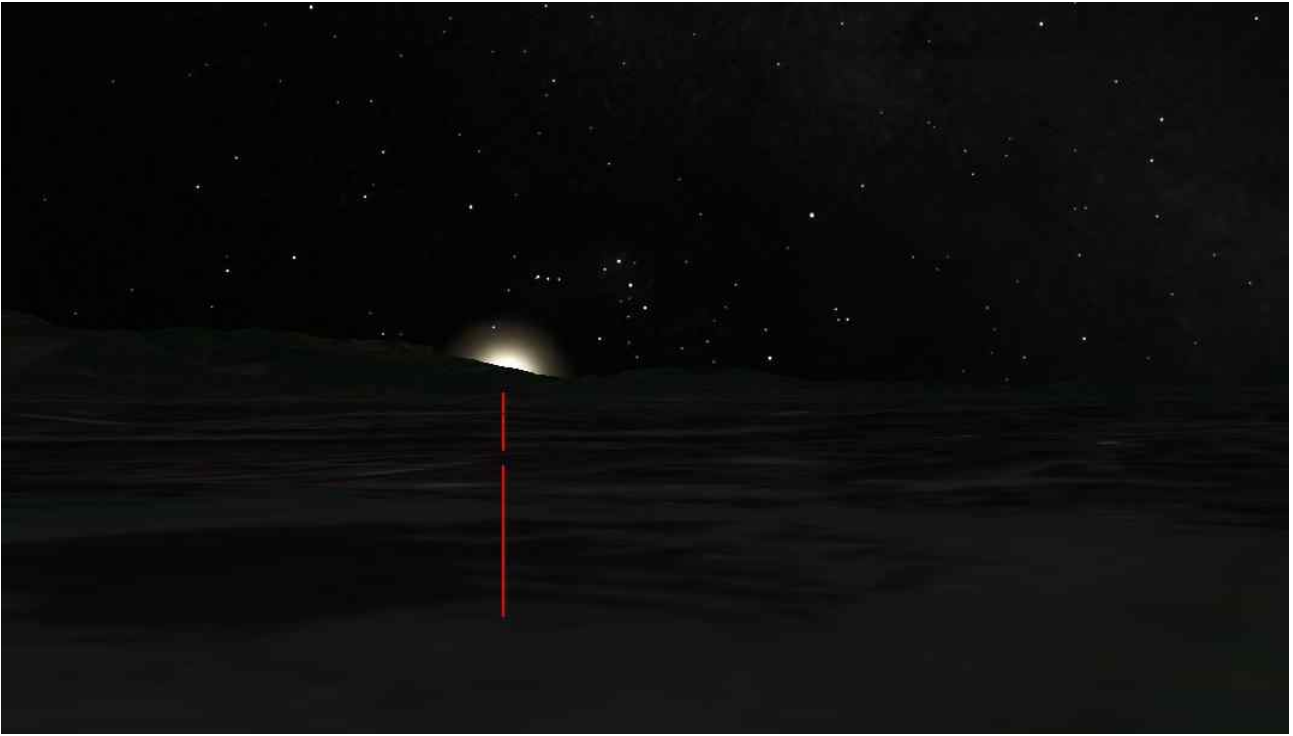

**Fig.3 Google Earth simulation of the sun setting along the Borobodur axis on a day of zenith passage.**

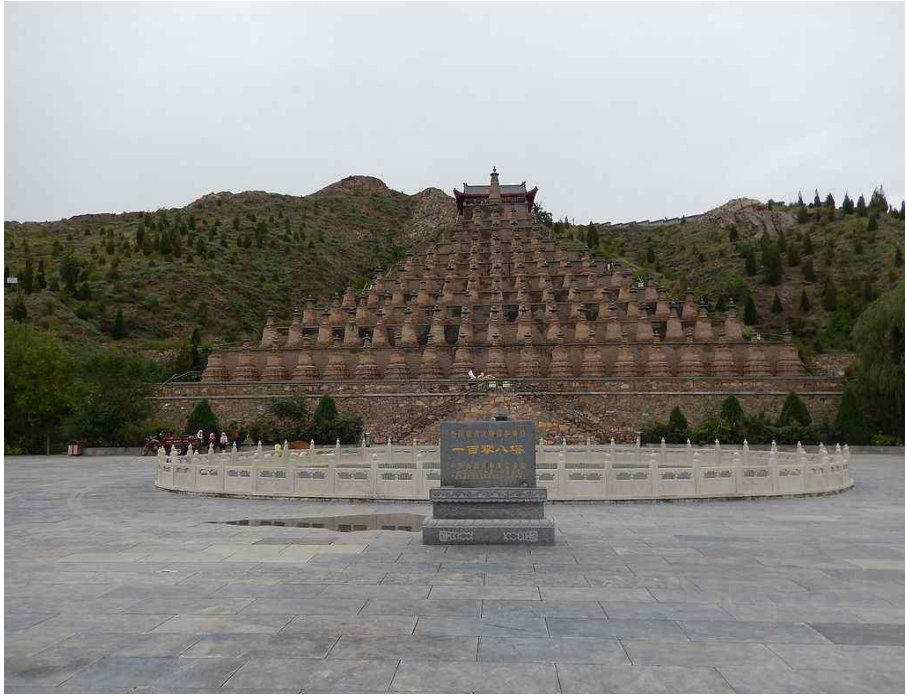

**Fig. 4 The 108 stupa monument.**